# Diffusive Limits of the Master Equation in Inhomogeneous Media

F. Sattin[1], A. Bonato[2], L. Salasnich[2,3,4]

[1]*Consorzio RFX (CNR, ENEA, INFN, Università di Padova, Acciaierie Venete SpA), Corso Stati Uniti 4, 35127 Padova, Italy*

[2]*Dipartimento di Fisica e Astronomia "Galileo Galilei", Università di Padova, Via Marzolo 8, 35131 Padova, Italy*

[3]*Consorzio Nazionale Interuniversitario per le Scienze Fisiche della Materia, Unità di Padova, Via Marzolo 8, 35131 Padova, Italy*

[4]*Istituto Nazionale di Ottica (INO) del Consiglio Nazionale delle Ricerche (CNR), Via Nello Carrara 1, 50019 Sesto Fiorentino, Italy*

*Abstract*

Diffusion is the macroscopic manifestation of disordered molecular motion. Mathematically, diffusion equations are partial differential equations describing the fluid-like large-scale dynamics of parcels of molecules. Spatially inhomogeneous systems affect in a position-dependent way the average motion of molecules; thus, diffusion equations have to reflect somehow this fact within their structure. It is known since long that in this case an ambiguity arises: there are several ways of writing down diffusion equations containing space dependence within their parameters. These ways are all potentially valid but not equivalent, meaning that the different diffusion equations yield different solutions for the same data. The ambiguity can only be resolved at the microscopic level: a model for the stochastic dynamics of the individual molecules must be provided, and a well-defined diffusion equation then arises as the long-wavelength limit of this dynamics.

In this work we introduce and employ the integro-differential Master Equation (ME) as a tool for describing the microscopic dynamics. We show that is possible to provide a parameterized version of the ME, in terms of a single numerical parameter ($\alpha$), such that the different expressions for the diffusive fluxes are recovered at different values of $\alpha$. This work aims to fill a gap in the literature, where the ME was shown to deliver just one diffusive limit. In the second part of the paper some numerical computer models are introduced, both to support analytical considerations, and to extend the scope of the ME to more sophisticated scenarios, beyond the simplest α-parameterization.

*1. Introduction*

Let us attempt to describe the dynamics of some fluid, quantified by its concentration $n(x,t)$ evolving in time and space. For simplicity, throughout this work we will restrict ourselves to one space dimension. In the absence of sink and sources, $n$ fulfils the conservation law

$$\frac{\partial n}{\partial t} = -\frac{\partial j}{\partial x} \qquad (1)$$

The kind of dynamics involved depends upon the analytical expression of the flux $j$. In the case of a collective motion where all the fluid parcels drift with the same velocity $V$: $j = nV$. Superimposed to this motion, the single fluid particles may experience disordered independent movements: Diffusion is the name given to the macroscopic realization of the random movement of large numbers of particles in space. The first experimental investigations on diffusion date back to the first half of nineteenth century, prominently with Thomas Graham's studies on the mixing of gases and of salts in water [1,2], and with Robert Brown and his studies on the movement of small pollen particles in aqueous suspension which were to be explained later by Einstein and Smoluchowski [2,3]. Another fundamental advancement was brought forth around the middle of the same century by Adolf Fick, who unified the diffusion in fluids with the conduction of heat in solids, studied earlier by Joseph Fourier [2,4]. The Fick-Fourier's (FF) law relates the flux $j$ with the gradient of the diffusing density $n$ through:

$$j = -D\frac{\partial n}{\partial x} \qquad (2)$$

The coefficient $D$ compactly summarizes the effect of the interaction between the individual particles and the surrounding milieu. In principle, if inter-particle interactions are not negligible, it may depend from $n$, too, but we will not consider this possibility here. In thermodynamics, Eq. (2) is an instance of a linear relationship between the flux and a thermodynamic force (the gradient of the free energy).

Eq. (2) was derived by Fick as a purely empirical relation. Alternatively, one can tackle a formal approach: the motion of each individual particle is modelled as a sequence of uncorrelated jumps, an instance of a Markovian stochastic process. This mathematical formalization leads to describing the trajectory of each individual particle through a Langevin equation. Then, the average over a large ensemble of such particles leads to the Fokker-Planck (FP) Equation [5,6]:

$$\frac{\partial n}{\partial t} = -\frac{\partial j}{\partial x} \equiv \frac{\partial}{\partial x}\left[\frac{\partial(nD)}{\partial x} - nV\right] \qquad (3)$$

In this expression, we may identify the overall flux as done by two contributions, that we will label as diffusive ($-\frac{\partial(nD)}{\partial x}$) and convective ($n\ V$). Analogously, we may generalize Eq. (2) by adding a convective contribution:

$$j \equiv -D\frac{\partial n}{\partial x} + Vn \qquad (4)$$

So far, no consideration has been made of the homogeneity of the background. As a matter of fact, the law (2) was worked out in contexts where it was impossible to discriminate experimentally any consequence of the inhomogeneity of the medium upon the dynamics of particles; implicit in it is therefore the postulate of the homogeneity of the medium. In homogeneous systems, $D,V,$ must be constant, since they are postulated to be dependent by the properties of the medium alone. It is straightforward to acknowledge that, in this case:

1) The fluxes $j$ in eqns. (3) and (4) are identical
2) There is a clear-cut unambiguous definition of convective and diffusive fluxes.

Just like (2-4) were reached by experiment and by theory, the coefficients $D,V$ appearing therein may be regarded either as unknown quantities to be fixed by matching with experiments, or as known from some more fundamental theory.

Real systems, however, are often inhomogeneous. This prompts to consider the possibility that the parameters quantifying the strength of the interaction with the background, $D,V$, become position-dependent: $D = D(x), V = V(x)$. In this case, (3) and (4) are no longer identical although they are still closely related:

$$\frac{\partial n}{\partial t} = \frac{\partial}{\partial x}\left[\frac{\partial(nD)}{\partial x} - nV\right] = \frac{\partial}{\partial x}\left[D\frac{\partial n}{\partial x} + n\frac{\partial D}{\partial x} - nV\right] = \frac{\partial}{\partial x}\left[D\frac{\partial n}{\partial x} - nV'\right] \qquad (5)$$

Thus, the flux initially written, e.g., in the Fokker-Planck form (3) may ultimately be written in the Fick-Fourier one through a redefinition of the convective term:

$$V' = V - dD/dx. \qquad (6)$$

This interchangeability may lead to speculate that the question of which flux to use is actually devoid of relevance: it has been argued in literature that only the total flux must be given physical meaning, not the diffusive and convective contributions separately [7]. This is trivially true as long as $D$ and $V$ are seen as pure fitting coefficients. However, the terms entering Eq. (6) come from distinct sources. Following van Kampen, we divide them in "geometric" and "thermal" terms ([8,9]). The latter expression originally implied that diffusion was caused by thermal agitation of the molecules. Since the gas we are considering is not necessarily made of real molecules, we will be employing it here in a broader acceptation, meaning any mechanism that acts on the microscopic disordered motion of the particles and thus affect the diffusivity *D*. Geometric terms, conversely, refer to the biasing on particle motion caused by background geometry such as external forcing, ratchet effects,…, ordinarily on large scales. Therefore, although formally, on the basis of Eq. (6) a varying diffusivity is indistinguishable from a genuine convection, we may expect to be potentially able to discriminate between the two on the basis of the physical mechanisms acting on the system under consideration. In this work we will postulate that geometric effects are not involved. This removes the previous ambiguity. For clarity we rewrite here down the versions of Eqns. (3,4) that we will be considering from now on:

$$\frac{\partial n}{\partial t} == \frac{\partial}{\partial x}\left[D\frac{\partial n}{\partial x}\right] \quad (7)$$

$$\frac{\partial n}{\partial t} = \frac{\partial}{\partial x}\left[\frac{\partial(nD)}{\partial x}\right] \quad (8)$$

At this stage, Eqns. (7,8) are no longer equivalent: when one and the same *D* is inserted in each of them, the resulting solutions *n(x,t)* are different (for the same boundary and initial conditions, of course). Furthermore, some classes of solutions are specific to the one or the other of these equations. One example is provided by uphill transport, sometimes observed in high-temperature magnetized plasmas, where the (matter or heat) flux goes along the same direction of the local gradient. This phenomenon is unexplainable within Eq. (7), since there flux flows only opposite to the gradient. Hence, in order to cope with this evidence, one has to enlarge the scope of Eq. (7) by allowing for some convection to exist, thereby reverting to Eq. (4), or equivalently to (8) [10] [For completeness we mention that it has been speculated also that other kind of transport might be at work here, producing Lévy flights, and therefore somewhat outside of the diffusive-convective paradigm].

The question: which expression between (7) and (8) is to be used, acquires therefore importance in this framework. The recent paper [11] highlights its relevance in the context of the modelling of ecological systems, assessing to which extent different choices for the diffusive fluxes do affect the results of several predefined test problems. Thus, it would be very convenient to have some guidelines telling us in advance of our analysis whether, in the system under consideration, it is more appropriate to use the one rather than the other equation. These guidelines have to be extracted from some knowledge of the microscopic physics driving the systems, and are accordingly expected to potentially be largely system-specific and non-universal. It is instead possible to answer on quite general grounds a weaker question: it is possible to establish a simple parametrization such that a single numerical parameter uniquely identifies whether a generic system is driven by the one or the other equation (7,8). This numerical parameter has a close connection with the microscopic stochastic dynamics, hence it is possible to assess its value only after a thorough knowledge of the system's microphysics.

There are several ways of introducing this formalism. One way is from the Lagrangian point of view, as adopted, e.g., by Lançon *et al* [12]. They modelled the motion of each individual particle as a Brownian motion with space dependent diffusion coefficient. An ambiguity arises in this case. Namely, the rule for updating the walker position writes:

$$x(t + dt) = x(t) + \eta\sqrt{2Ddt} \qquad (9)$$

Where η is a white noise of unit amplitude. However, the question arises: at which location *D(x)* should be evaluated? At the initial position, *x(t)*; at the final one, *x(t+dt)*, or somewhere else ? The need of making such a choice goes under the name of Ito-Stratonovich convention. Adhering to the one or the other choice ultimately leads to Eq (7) or (8) when one goes from the single particle to the fluid-like picture of the motion. Details can be found in Lançon *et al*.

An Eulerian viewpoint is possible as well. This time the location of space is held fixed, and accounting is done of fluid elements entering and leaving it according to given stochastic rules. The resulting balance equation is named *Master Equation* as introduced by Van Kampen [13]. The Master Equation (ME) formalism and the Browian walker one are, of course, different ways of expressing the same physics (although the former one is more flexible) and therefore are expected to produce the same results. To the best of our knowledge, however, the exercise of extracting diffusion equations (7,8) from the ME has been fulfilled so far only partially, by van Milligen *et al*, who recovered Eq. (8) [14]. No analogous result for (7) has been produced. This led van Milligen *et al* to argue that perhaps Eq. (7) does lack a true microscopic justification. In this paper, instead, we will complete this exercise by showing how Eq. (7) can be extracted from the ME in analogy with the Brownian walker result. Section 2 provides a brief introduction to the ME. Sections 3 and 4 are devoted to show how Eqns. (8), (7) respectively do arise as suitable limits of the ME. Section 5 supplements the previous analytical results with numerical exercises: lattice models are designed and implemented as computer codes and we will show that, on the basis of the stochastic rules imposed, the numerical simulations do conform to the predictions done in the previous sections. Indeed, we have remarked earlier that the ME is fairly a flexible tool for the investigation of stochastic system. Still in section 5 we will validate this claim, by producing an instance of a system whose dynamics is modelled by the ME and that cannot be reduced to either Eq. (7) or (8). Finally, section 6 provides a brief recap of the results.

*2. The Master Equation*

The ME is an integro-differential equation expressing the rate of change for the local scalar density *n* in terms of transition probabilities. In one spatial dimension and with consideration of just one transported quantity it writes

$$\frac{\partial n(x,t)}{\partial t} = -\frac{n(x,t)}{\tau} + \int dz\, n(z,t) \frac{P(x,z)}{\tau} \qquad (10)$$

In (10), *P(x,z)* represents the probability for the lump of matter *dz n(z)* to be moved from *z* to *x*, and $\tau$ sets the corresponding temporal scale, which may depend upon *x* as well. Thus the first term in the r.h.s. of (10) represents the rate of particles leaving the location *x*, whereas the integral stands for the total rate of particles that, started elsewhere, land on *x*. Eq. (10) does not account for the presence of any source and/or sink; we will not consider them in this work although they can be added straightforwardly. We cursorily note that *P* may be considered as a transfer function without necessarily a connection with probability. A deterministic version of Eq. (10) is used, e.g., in biophysics, where it is known as Amari's equation [15].

The limits of integration in Eq. (10) are left unspecified. The reader may assume that they range from $-\infty$ to $+\infty$; we will not address here such issues as finite-size boundary effects (some related analysis can be found in [16]).

Eq. (10) may contain virtually all the information available about the system, except for the velocity degrees of freedom of the particles. It is is fairly flexible, too: depending on the functional form chosen for *P*, $\tau$, it can accommodate a wide range of dynamics, from sub- to super-diffusive transport, including possibly nonlinearities. Indeed, in the form (10) it has already been circumscribed from a more general expression (Generalized Master Equation) [17] that involves a convolution not only over space, but over time as well: $P = P(x, z, t, t')$. Explicit consideration of a finite temporal memory may lead to the emergence of a vastly more complicated dynamics, varying from diffusive to wave-like, as argued by Maxwell, Cattaneo and Vernotte [18].

*3. From the ME to the Fokker-Planck flux*

In most practical applications, one has available just coarse-grained information; i.e., not knowledge of the full $P,\tau$, but just averages $<P>, <\tau>$ over finite regions of the system. Mathematically, this amounts to saying that only the long-wavelength limit of Eq. (10) is actually relevant. In this and the next section we will derive the result that by removing small-scale details—i.e. taking its long-wavelength limit—Eq. (10) reduces a parabolic partial differential equation, the diffusion equation, and that can be either in the form (7) or (8).

For simplicity, throughout this work, we will drop any dependence of τ from space or time: it will just play the role of a constant, characteristic time scale of the process. Accordingly, the whole physical content of the theory is brought by *P*. Since we are dealing with position-dependent systems, one could legitimately wonder whether dropping any spatial dependence from τ does impose too severe constraints upon the dynamics we may model. Setting τ = τ(x) does—of course—affect quantitatively the shape of the resulting equations, since additional terms proportional to τ'(x), τ''(x) do appear. However, conclusions drawn in the next sections will remain qualitatively unaffected.

In order to carry out any further analysis one needs first to make explicit the dependence of *P* upon its arguments. In spatially homogeneous systems, the only dependence can enter through the jump length *(x-z)* since *P* is invariant under translations. In inhomogeneous systems, instead, this invariance must be broken, and explicit dependence of *P* upon *x, z* separately must appear. The most intuitive choice is to make *P* dependent from the starting location as well:

$$P(x, z) \equiv P(x - z; z) \tag{11}$$

Heuristically, this choice is close to the standard way of looking at problems in dynamics—or in computer programming: the evolution of a system is determined (although, in this case, only on a statistical basis) by the law of motion (*P*) and by the initial conditions (*z*).

It is convenient switching to the variable $\Delta = z - x$ in Eq. (10):

$$\frac{\partial n(x,t)}{\partial t} = -\frac{n(x,t)}{\tau} + \frac{1}{\tau}\int d\Delta\, n(x+\Delta,t) P(-\Delta; x+\Delta) \qquad (12)$$

The long-wavelength limit is taken by postulating that the jump probability *P* is practically zero beyond some maximum jump length $\Delta_{max}$, and that both *n* and *P* vary slowly over lengths of order $\Delta_{max}$. Therefore, the argument of the integral in (12) is expanded around *x* in powers of $\Delta$, and the Taylor expansion is stopped to some finite order. Pawula theorem [5] warrants that, by stopping to the second order, no such unphysical artefacts as locally negative densities may occur. The true justification for the truncation to the second order, however, has been provided only quite recently by Ryskin [19]. Ryskin's proof is essentially a consequence of the Central Limit Theorem, and states that for analytic *P*'s and over time scales just moderately longer than τ, the dynamics always converges to be diffusive. Since Ryskin's result is essential for this paper, we will provide in the Appendix a sketch of his proof.

Armed with these results we eventually carry out the expansion of (12):

$$\frac{\partial n(x,t)}{\partial t} = -\frac{n(x,t)}{\tau} + \frac{1}{\tau}\int d\Delta\, nP + \frac{1}{\tau}\int d\Delta\, \Delta \frac{\partial}{\partial x}(nP) + \frac{1}{\tau}\int d\Delta\, \frac{\Delta^2}{2}\frac{\partial^2}{\partial x^2}(nP) \qquad (13)$$

In Eq. (13), for brevity, we have not written explicitly the arguments of *n* and *P*: *n* = *n(x,t)*, *P* = *P(-Δ,x)*.

The two first terms in the r.h.s. of (13) mutually cancel by virtue of the normalization

$$\int d\Delta\, P(-\Delta; x) = 1 \qquad (14)$$

Eq. (13) takes thus the form of a conservation equation:

$$\frac{\partial n}{\partial t} = \frac{\partial^2(D(x)n)}{\partial x^2} - \frac{\partial(U(x)n)}{\partial x} \qquad (15)$$

with

$$U(x) = \frac{1}{\tau}\int d\Delta\, \Delta\, P(-\Delta; x),\quad D(x) = \frac{1}{\tau}\int d\Delta\, \frac{\Delta^2}{2} P(-\Delta; x) \qquad (16)$$

In accordance with the guidelines set up in the Introduction, we will restrict to scenarios without convective flux. This is achieved by imposing unbiased jump probability: $P(\Delta;x) = P(-\Delta;x)$ [1], thus $U(x) = 0$, and

$$\frac{\partial n}{\partial t} = \frac{\partial^2 (D(x)n)}{\partial x^2} \qquad (17)$$

Eq. (17) is thus a continuity equation for $n$, the flux being written in the FP form (8). It comes fairly straightforwardly from the postulate (11), which looks like quite natural.

This result thus may suggest that the FP is the natural long-wavelength limit of the ME—within the above constraint of unbiased *P*. Van Milligen *et al* [14] supported this conclusion both by recalling earlier numerical results—namely, the computer simulations of particles hopping on a lattice by Collins *et al* [20]—as well as carrying out and modelling some experiments of viscous fluids dynamics [21].

We mention that this kind of question (which choice is physically motivated) has often occurred in the literature [5,22]. We have repeatedly stated throughout this work that the question has no answer, since both choices may be valid. Actually, unambiguous evidence for Fickian transport in inhomogeneous media with unbiased *P does exist* in literature. For instance, one may mention (*i*) generic one-dimensional Hamiltonian systems [23]; (*ii*) the experiments on the dynamics of colloidal particles [12]; (*iii*) the analytical and computer models of particles scattering against a Lorentz background made by Bringuier [9], as well as (*iv*) the extensive study by Schnitzer [24]. Several of these results are based upon microscopic models of the stochastic dynamics, thus there is no doubt that the Fick's flux is a valid limit for some classes of systems.

*4. From the ME to the Fick-Fourier flux*

It is therefore clear that we have not yet exhausted all the physics potentially embedded into Eq. (10). The purpose of this section is to make explicit the constraints imposed for the derivation of Eq. (17) and see how they have to be relaxed to allow for wider classes of solutions. To this purpose, we will outline a recipe patterning the classical Itô-Stratonovich [5,6,25] one within the framework of the Master Equation. Let us start by looking at the structure of the arguments of the transition probability $P$ within a particle-hopping picture: a generic particle is bound to location $z$ for a duration $\tau$, and then is propelled away. The initial conditions of the particle are completely fixed (although only in a statistical sense) by the local background at the starting point $z$ (whence the appearance of $z$ as second argument of $P$ in Eq. 11), whereas during its travel it suffers an interaction with its environment that damps its motion up to a complete stop at point $x$, whence the meaning of the first argument of $P$ as the total length travelled: $x - z$. This picture is fairly intuitive, and for some classes of systems it provides a close approximation of the real dynamics, but we should remember that Eq. (10) is meant generically to give just a coarse description for the actual dynamics, which is much more complicated since involves kinetic degrees of freedom, too: we are neglecting, for instance, all effects related to the finite inertia of the moving particles. It is the same kind of ignorance about the true path travelled by the particle that, in Brownian motion (Eq. 9), leads to choosing between the different Itô or Stratonovich recipes. Therefore we should expect that $P$ depends explicitly not only on the starting location $z$ but potentially all the points between $z$ and the final one, $x$. It is intuitive that such a detailed accounting would likely lead to a fairly complicated expression when inserted into Eq. (10). However, since only the initial ($z$) and the final ($x$) locations are well defined, not the exact trajectory between the two a more reasonable argument is that just them should appear as arguments of $P$. The simplest ansatz is to allow for a linear combination of the two:

$$P = P(x-z; \bar{x}), \bar{x} = (1-\alpha)z + \alpha x, 0 \leq \alpha < 1 \tag{18}$$

---

[1] We add a comment in order to avoid potential confusion about this regard. The complete transition probability *P(x-z;x;z)* may be—and usually *is* in inhomogeneous environments—biased: $\int d\Delta\, \Delta\, P \neq 0$, but the average defining *U* in Eq. (16) involves the reduced probability *P(x-z';x;z=x)* in which *z* is set equal to *x* in all arguments but in the step length. It is this reduced probability that has to be unbiased.

Thus, with $\Delta = z - x$ and $G = P/\tau$ the ME writes

$$\frac{\partial n(x,t)}{\partial t} = -\frac{n(x,t)}{\tau} + \int d\Delta\, n(x+\Delta,t) G(-\Delta, x+(1-\alpha)\Delta) \tag{19}$$

Now we expand the terms inside the integral in power series around $x$ dropping as customary all terms involving odd powers of $\Delta$. In order to keep expressions short, in the next lines we label the spatial derivative with the apex sign: $' \equiv \partial/\partial x$ and the time derivative as $\frac{\partial}{\partial t} \to \partial_t$.

We get

$$\partial_t n = \int d\Delta \frac{\Delta^2}{2}\left\{2(1-\alpha)G'n'+Gn''+(1-\alpha)^2 G''n\right\}+\int d\Delta\{nG\}-\frac{n}{\tau} \tag{20}$$

By replacing the integrals in (20) using *D* given in (16) we get

$$\partial_t n = \{Dn' + D'(1-2\alpha)n\}' + \left[\alpha^2 D''n + \int d\Delta\{nG\} - \frac{n}{\tau}\right] \tag{21}$$

The normalization condition for particles jumping from the fixed location *z* to an arbitrary one *x* reads

$$\int dx\, P(z,x) = 1$$

After inserting Eq. (18) for *P*:

$$1 = \int dx\, P[x-z;(1-\alpha)z+\alpha x] \tag{22}$$

We rewrite in this expression the arguments of *P* as $P = P(-\Delta; z-\alpha\Delta)$, and expand (22) up to the second order in $\Delta$:

$$1 = \int d\Delta\, P[-\Delta;z] - \alpha \frac{d}{dz}\int d\Delta\, \Delta P[-\Delta;z] + \alpha^2 \frac{d^2}{dz^2}\int d\Delta\, \frac{\Delta^2}{2} P[-\Delta;z] + \ldots$$

The first-order term in α vanishes by virtue of the symmetry of *P*. After the multiplication by $n(z)/\tau$ and the replacement of the second integral with *D* we get

$$\frac{n}{\tau} = \int d\Delta\{nG\} + \alpha^2 D''n \tag{23}$$

Therefore the term inside square brackets in Eq. (21) vanishes and we are left with

$$\partial_t n = \{Dn' + D'(1-2\alpha)n\}' \tag{24}$$

Let us now specialize Eq. (24): first consider the case $\alpha = 0$. We get accordingly

$$\partial_t n = (nD)'' \tag{25}$$

which is just the FPE once again, as expected since Eq. (18) reduces now to (11). The choice $\alpha = 1/2$ yields instead:

$$\partial_t n = (Dn')' \tag{26}$$

*i.e*, the FF diffusion.

We can thus conclude that the ME may actually produce Fickian transport as long as its diffusive limit makes sense.

For completeness, let us add some brief comments about the third special case $\alpha = 1$.

$$\partial_t n = (Dn' - D'n)' \tag{27}$$

It is interesting to perform a comparison between (25) and (27). Both are instances of "flux without a gradient", i.e., fluxes can be sustained even in the absence of a gradient in the concentration *n*. By setting *n'* = 0 in both we get respectively

$$\partial_t n = nD'' \qquad \partial_t n = -nD'' \qquad (28)$$

Hence, the two cases ($\alpha = 0,1$) are related by the "mirror" symmetry $D'' \leftrightarrow -D''$.

We conclude this section with a minor comment about the parallels between our procedure and that of Lançon *et al* [12] using the Brownian walk formalism. We point out how the ME approach grants a small advantage: namely, their derivation could retain terms containing *D*, *D'* but not higher derivatives (unlike ours) since it would turn the Brownian walker algorithm updating the position ($x(t)$ → $x(t+dt)$) into an algebraic equation for $x(t+dt)$ of order larger than one, with related issues of multiple roots.

*5. Numerical experiments: Lattice models of diffusion*

The ansatz (18) suggests that *final* conditions are to be involved in order to recover the FF flux. One might wonder which systems do fulfil it. In this section we design two different models for particles hopping between nodes of a one-dimensional lattice, hence variants of the Collins' model [20]. Differences are in the statistical rules obeyed by the particles. Model #1 provides a flexible knob easily interpolating between different dynamics. Model #2 is somehow less amenable to direct inspection of its emergent dynamics, but implements perhaps a more realistic mechanism.

*5.1 Model #1*

At each time step, particles perform a random jump from their starting node to some neighbouring one. The width of the jump, *i.e.* the number of nodes the particle can travel, is picked randomly from a uniform distribution $[-\ell(i), \ell(i)]$, where *i* is the starting node, and $\ell$ a maximum jump length that depends on the starting location. Furthermore, each node *j* is assigned an *acceptance rate* $0 \leq P_a(j) \leq 1$. Once a particle has been chosen to hop from node *i* to node *j*, a second statistical test is performed based upon the acceptance rate: it succeeds with probability $P_a(j)$, and the particle moves from *i* to *j*, whereas with probability $1 - P_a(j)$ the test fails, and in this case the particle does not move. The hopping probability thus writes

$$P(j|i) = P_j(j - i|i) \times P_a(j) \qquad (29)$$

Provided that $\ell$ is small with respect to the length of the lattice, we can consider a continuous version of the ME (where we set $\tau = 1$)

$$\frac{\partial n(x,t)}{\partial t} = -n(x,t) + \int d\Delta\, n(x+\Delta,t) P_j(-\Delta, x+\Delta) P_a(x) \tag{30}$$

which, after the usual Taylor expansion, writes

$$\begin{aligned}\frac{\partial n(x,t)}{\partial t} &= -n(x,t) + \int d\Delta\, n(x,t) P_j(-\Delta,x) P_a(x) + \int d\Delta\, \Delta P_a(x) \frac{\partial}{\partial x}\bigl(n(x,t) P_j(-\Delta,x)\bigr) \\ &\quad + \frac{1}{2}\int d\Delta\, \Delta^2 P_a(x) \frac{\partial^2}{\partial x^2}\bigl(n(x,t) P_j(-\Delta,x)\bigr) \\ &= P_a(x)\left[-\frac{\partial}{\partial x}\bigl(n(x,t) U_j\bigr) + \frac{\partial^2}{\partial x^2}\bigl(n(x,t) D_j\bigr)\right]\end{aligned} \tag{31}$$

$$U_j = -\int d\Delta\, \Delta P_j(-\Delta,x) = 0, \quad D_j = \frac{1}{2}\int d\Delta\, \Delta^2 P_j(-\Delta,x)$$

The odd moment $U_j$ vanishes because $P_j$ has mirror symmetry and (31) reduces to

$$\frac{\partial n(x,t)}{\partial t} = P_a(x) \frac{\partial^2}{\partial x^2}\bigl(n(x,t) D_j\bigr) \tag{32}$$

Eq. (32) is obviously not in the form of a Fokker-Planck Equation—unless $P_a$ is a constant--but can be cast into it by defining the auxiliary functions $v = \frac{n}{P_a}$, $D = P_a D_j$:

$$\frac{\partial v}{\partial t} = \frac{d^2}{dx^2}(v\, D)$$

We check now that the presence of the $P_a$ function outside the derivative term in (32) does not cause problems with the particle conservation. We postulate $P_a(x)$ to be a smooth function. To lowest order, we can get $P_a$ constant, which obviously turns Eq. (32) into a continuity equation for $n$. Then, we go to the next order by linearly expanding $P_a$ around some fixed point, conventionally $x = 0$: $P_a(x) = P_a(0) + P_a{'}(0)\, x$. This allows rewriting Eq. (32):

$$\frac{\partial n(x,t)}{\partial t} = \frac{\partial}{\partial x}\left[P_a(x)\frac{\partial(nD_j)}{\partial x} - \frac{dP_a}{dx} n D_j\right] = \frac{\partial}{\partial x}\left[\frac{\partial}{\partial x}(Dn) - 2nD\frac{d\ln P_a}{dx}\right] \tag{33}$$

which is still in conservative form.

The convective flux is still geometric in nature, since is proportional to space derivatives of the jumping probabilities, but it has a more elaborated dependence than just from $dD/dx$, hence cannot be derived from within the Itô-Stratonovich α-parametrization.

Eq. (33) reduces to the FPE when $dP_a/dx = 0$. Conversely, if $P_a$ is not constant, the probability for a particle to jump between two nodes depends on the arrival site as well as the departure one, and we expect non-FP features to arise. Fick's flux arises in connection with the condition:

$$\frac{d \ln D}{dx} = 2 \frac{d \ln P_a}{dx} \rightarrow D_j \propto P_a \tag{34}$$

This, together with $D_j(x) = \frac{1}{2}\langle \ell^2 \rangle = \frac{1}{6}\ell^2(x)$, that comes from our choice of the statistical distribution, establishes that Fick's diffusion arises for just a specific functional form of the jumping length:

$$\ell \propto \sqrt{P_a}. \tag{35}$$

In the following we show the result of a numerical simulation done tracking the evolution of $N_p = 10^6$ particles over a lattice with $N = 2048$ nodes: a Monte Carlo implementation of solution of the Master Equation. All particles are initially located at $i = N/2$ and reflecting boundaries are imposed at both sides. As far as the acceptance rate is concerned, we consider two scenarios:

(A) variable acceptance rate: $P_a = 0.1 + 0.85(i/N)$

(B) constant acceptance rate: $P_a \equiv 1 \quad \forall i$

The jump length is instead taken in both cases as $\ell = \left\lfloor 12\sqrt{0.1 + 0.85(i/N)} \right\rfloor$ (where […] means the integer part); thus it fulfils constraint (35) in the scenario (A), whereas yields Fokker-Planck flux in scenario (B). There is nothing special about these numerical values; they have simply been chosen on the basis of the following—fairly trivial—considerations: (i) the larger $\Delta\ell/\Delta i$, the more any effect related to inhomogeneity shows up. However, $\ell(i)$ must be small enough in order for $D$ to make sense as a local quantity: hops must not be too large. With the choice above we get that $D_{max} > 100\, D_{min}$ and $\ell_{max}/N \approx 1/200$ which fulfil both these constraints. (ii) The same rationale applies to $P_a$ as well: the larger $dP_a/dx$, the more clearly the departure from FP flux appears, but $P_a$ must stay in the range $0 < P_a \leq 1$, and values too close to zero make the numerical treatment cumbersome, since particles take very long times to sample accurately these regions.

We compare the particle simulation with the numerical solution of the diffusion equation for both the choices Fokker-Planck (Eq. 8) and Fick-Fourier (Eq. 7). In the figure the dots are the number of particles found at the different nodes after $t = 10^4$ time steps, the black (red) curves the FP and FF solutions respectively. There is a clear agreement between expectations and numerical results in both cases.

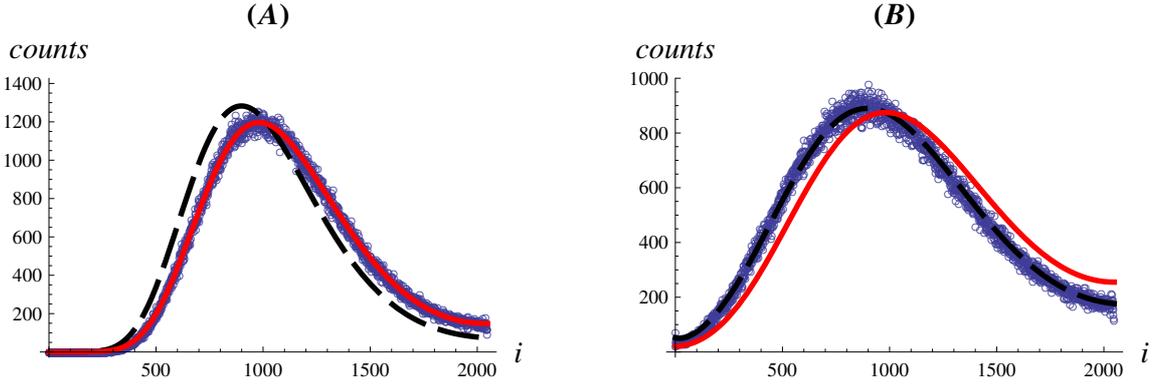

**Fig. 1.** Left plot, particle distribution after $10^4$ steps for scenario (A) with variable acceptance rate; right plot, the same for scenario (B) with constant acceptance rate. Dots are from the numerical simulation of the particle model; red solid curve is the solution of fluid diffusion equation using FF's flux (Eq. 7); black dashed curve, the same using FP flux (Eq. 8).

*5.2 Model #2*

In model #1 the particle is required to pass a test relative to the destination node *before* it actually reaches the node itself. This time, we restore the locality in the model: all the tests are taken at the node occupied by the particle.

We consider a lattice version of mono-energetic particle motion: at each time clock particles are moved of exactly one node: say ($j$-$1$ → $j$). However, *after the displacement is done* a test is carried out: with probability $1- q(j)$ the particle conserves its direction during the next time clock, whereas with probability $q(j)$ it will reverse its direction. The next step, in the two cases, will be either ($j$ → $j+1$) or ($j$ → $j-1$). Thus, the total jump of the particle is defined by summing the number of nodes $n$ travelled between two successive changes of direction, its probability being $P = q(j) \prod_{n-1} (1 - q(l))$, where the index $l$ stands for the nodes' indices between the starting and the final one; therefore $P$ will depend on all the nodes visited as well. This way, the appearance of the final location between the arguments of $P$ appears clearer: it is not necessary that the particle collects information about the arrival site in advance of its hopping, like in model#1. The model built this way is a version of random walk, thus its large-scale dynamics has to be diffusive. Unlike model #1, in this case, we were not able to guess an explicit expression for the diffusivity; however, since $P$ is position-dependent, $D$ must be as well, $D = D(x)$-

We have carried out a numerical Monte Carlo simulation for this system, which yields the solution of the Master Equation. Starting from a collection of particles all placed exactly in the middle of the lattice, we have let them to move randomly over times so long that the final stationary equilibrium is reached. The equilibrium density is shown in Fig. (2), and is spatially constant. Furthermore, at equilibrium, the flux must be null. We were not able to work out explicitly an expression for the diffusivity $D$, but we know that the stationary solution using the FP flux is $-d(Dn)/dx = 0 \rightarrow n \propto 1/D \neq$ constant. Hence, regardless of the value of $D(x)$, the flux cannot be in the Fokker-Planck form. Conversely, the FF flux is $-Ddn/dx$, and is consistent with the stationary solution $n =$ constant.

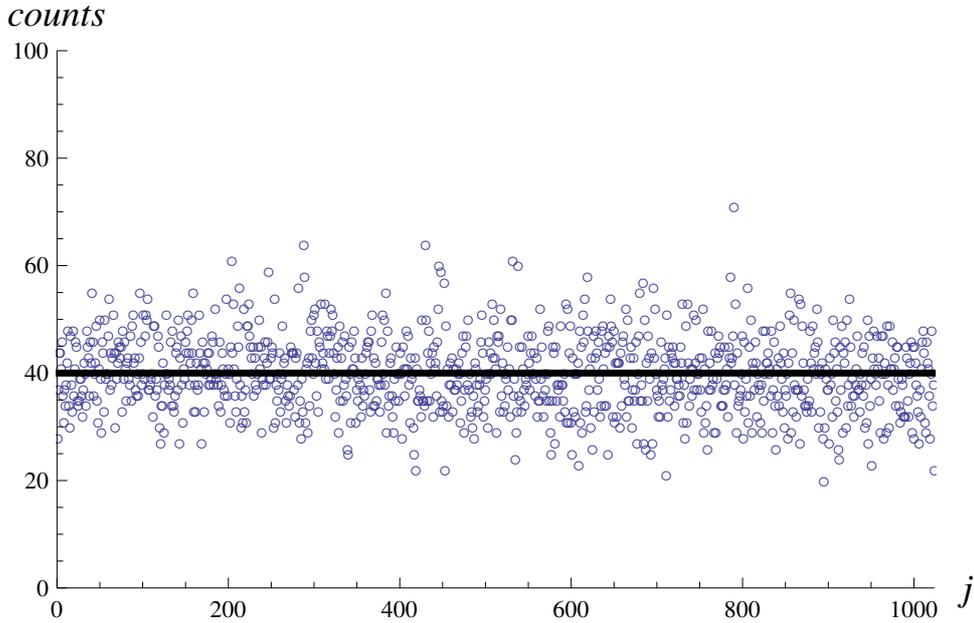

**Fig. 2.** Dots: particle distribution for the model #2 after $2.5 \times 10^5$ time steps. Parameters of the simulation are: number of nodes $N = 1024$; number of particles $N_p = 40 \times 1024$; particles initialized at $j = N/2$; $q(j) = 0.1 + 0.4*j/N$. Reflecting boundaries are used. The resulting final distribution is uniform within the statistical noise: for reference, the black line is the perfectly constant density. This result is at a variance with expectations from Fokker-Planck flux, whereas it is consistent with Fick-Fourier flux.

## 6. Conclusions

In this paper we have provided a simple procedure to derive diffusion equations as several different limits from within the single framework of the ME. We have worked out an analytical expression, Eq. (24), that can interpolate between both Fick-Fourier and Fokker-Planck limits in dependence of a single numerical parameter $\alpha$. We have highlighted that the value of $\alpha$ should arise from the knowledge of the microscopic physics of the system examined. In most actual situations, it is likely difficult to extract it, but we have already provided several instances where this exercise was carried out [9,20]. Other examples include [26] and—fairly recently—[27], although these results have been latter questioned [28]. Regardless of the fact that one is able to determine *a priori* which is the best expression for the diffusive flux to be employed, the main conclusion stressed in this work is that both choices are legitimate long-wavelength limits of some underlying microscopic model. Furthermore, we have shown that the ME formalism is much more comprehensive than the Langevin equation one. The analytical conclusions have further been supported by implementing and solving numerically simple intuitive models.

*Acknowledgments*

FS wishes to thank Prof. E. Bringuier and Prof. M. Baiesi for providing him with some of the literature quoted, Dr. D.F. Escande, Dr. S. Cappello, Dr. I. Predebon for reading drafts of this paper and Prof. G. Ryskin for interesting discussions about his proof. This project has received funding from the European Union's Horizon 2020 research and innovation programme under grant agreement number 633053. The views and opinions expressed herein do not necessarily reflect those of the European Commission. LS thanks MIUR for partial support (PRIN Project 2010LLKJBX).

*Appendix*

The rationale of Ryskin's result is based upon the Central Limit Theorem (CLT). The total displacement of a particle is approximated as the sum of several uncorrelated jumps. The single jumps are not necessarily identical, but picked up from some statistical distribution. Regardless of the details of the distribution, provided that the variance of the jumps remains finite, the CLT warrants that the total displacement distributes according to a statistical distribution that quickly approaches a Gaussian distribution after even a moderate numbers of steps.

For brevity we will sketch Ryskin's proof for homogeneous systems only. Its generalization to inhomogeneous systems adds some mathematical labour but does not differ conceptually.

In homogeneous systems $P$ depends just from the difference of its arguments: $P = P(x-z)$, as explained in Section 3. This allows for a dramatic simplification after the Fourier transform of Eq. (10) is taken:

$$\frac{\partial \tilde{n}(k)}{\partial t} = -\frac{\tilde{n}(k)}{\tau} + \frac{\tilde{n}(k)\tilde{P}(k)}{\tau} \tag{A1}$$

Eq. (A1) can be solved analytically:

$$\tilde{n}(k,t) = \tilde{n}(k, t=0) \times \exp\left(\int_0^t \frac{dt}{\tau}(\tilde{P}(k) - 1)\right) = \exp\left(\frac{t}{\tau}(\tilde{P}(k) - 1)\right) \tag{A2}$$

Formally, $n(x,t)$ comes from the inverse transform of (A2) and it does not appear analytically computable for generic $P$. However, we can write

$$\tilde{P}(k) = \int dz\, e^{ikz} P(z) = \sum_{m=0}^{\infty} \frac{(ik)^m}{m!} \int dz\, z^m P(z) \tag{A3}$$

We will be considering just specularly symmetric transitions, as customary: $P(x-z) = P(z-x)$; the moments in Eq. (A3) become

$$\int dz\, P = 1 = \tilde{P}(k=0) \tag{A4}$$

$$\int dz\, z\, P = 0 \tag{A5}$$

$$\int dz\, z^2\, P = \sigma^2 \tag{A6}$$

$$\int dz\, z^m\, P = <z^m> \equiv \mu_m \quad m \geq 3 \tag{A7}$$

Let us now rewrite Eq. (A3) using the trigonometric expression of the exponential:

$$\tilde{P}(k) = \int dz \cos(k\,z) P(z) + i \int dz \sin(k\,z) P(z)$$

$$\rightarrow |\tilde{P}(k)|^2 = \left(\int dz \cos(k\,z) P(z)\right)^2 + \left(\int dz \sin(k\,z) P(z)\right)^2$$

$$= <\cos(k\,z)>^2 + <\sin(k\,z)>^2 \tag{A8}$$

This implies that $|\tilde{P}(k)| < 1$ for generic $k \neq 0$, and

$$|\tilde{P}(k)| \rightarrow 0,\, k \rightarrow \infty. \tag{A9}$$

To demonstrate this, let us note that $sin(k\,z)$, $cos(k\,z)$ are periodic with wavelength $\lambda = \frac{2\pi}{k} \rightarrow 0, k \rightarrow \infty$, while $P$ is a smoothly varying function, hence is almost constant over $\lambda$: $P(z) \approx P(z + \lambda) = P_0$. Therefore

$$\int_0^\lambda dz \cos(k\,z) P(z) \approx P_0 \int_0^\lambda dz \cos(k\,z) \approx 0 \tag{A10}$$

(the same holds for $sin(k\,z)$).

We define $m$, $\Delta t$ such that $t = j\,\Delta t$, with $j$ integer and $\Delta t \approx O(\tau)$. Eq. (A2) becomes

$$\exp\left(\frac{t}{\tau}(\tilde{P}(k)-1)\right) = \left[\exp\left((\tilde{P}(k)-1)\frac{\Delta t}{\tau}\right)\right]^j =$$

$$\left[\exp\left(\frac{\Delta t}{\tau}\left(-\frac{k^2\sigma^2}{2}+\sum_{m=3}^{\infty}\frac{(ik)^m}{m!}\mu_m\right)\right)\right]^j \tag{A11}$$

In the last line of (A11) we have taken advantage of (A4-A7).

Let us define $\xi = j^{1/2}k$. Eq. (A11) becomes

$$\frac{\tilde{n}(k,t)}{\tilde{n}(k,t=0)} = \exp\left[\frac{\Delta t}{\tau}\left(-\frac{\xi^2\sigma^2}{2}+\sum_{m=3}^{\infty}\frac{1}{j^{\frac{m}{2}-1}}\frac{(i\xi)^m}{m!}\mu_m\right)\right] \tag{A12}$$

Then, we consider separately the two limits $\xi>1$ and $\xi\leq 1$[2]. The former limit corresponds, for any fixed time, to taking $k \to \infty$ and therefore the result (A9) holds: there is not contribution to the density from features at these wavelengths. Conversely, when $\xi\leq 1$ the first term inside the exponent in Eq. (A12) dominates over the others hence we can retain just it and, reverting to the original variables

$$\tilde{n}(k,t) = \tilde{n}(k,t=0) \times \exp\left[-\frac{t}{\tau}\frac{(k\sigma)^2}{2}\right] \tag{A13}$$

which is the propagator of the diffusion equation, with diffusivity $=\frac{\sigma^2}{2\tau}$. This concludes the proof.

---

[2] Notice that the boundary between $\xi>1$ and $\xi\leq 1$ is a dynamical one: the range of $k$'s contributing to (A13) does vary with time (*i.e.* with $j$).